\documentclass[12pt, letterpaper]{article}
\usepackage[utf8]{inputenc}

\title{Causes of jets in the quasi-perpendicular magnetosheath}
\author{Primo\v{z} Kajdi\v{c}$^1$, Savvas Raptis$^2$, X\'ochitl Blanco-Cano$^1$, Tomas Karlsson$^2$}

\usepackage{graphicx}
\usepackage{amsmath}
\usepackage{amssymb}
\begin{document}

\maketitle
\begin{center}
$^1$Departamento de Ciencias Espaciales, Instituto de Geof\' isica, Universidad Nacional Aut\'onoma de M\'exico, Ciudad Universitaria, Ciudad de M\'exico, Mexico, $^2$Space and Plasma Physics, KTH Royal Institute of Technology, 10405, Stockholm, Sweden
\end{center}

\begin{itemize}
\item Jets in the quasi-perpendicular (Qper) magnetosheath may be of different origin than those in the quasi-parallel magnetosheath.
\item Some casuses of jets in the Qper magnetosheath are known phenomena, such as mirror mode waves, current sheets and reconnection exhausts.
\item Some jets in the Qper magnetosheath are due to passing magnetic flux tubes, downstream equivalents of travelling foreshocks.
\end{itemize}

\begin{abstract}
Magnetosheath jets are currently an important topic in the field of magnetosheath physics. It is thought that 97~\% of the jets are produced by the shock rippling at quasi-parallel shocks. Recently, large statistical studies of magnetosheath jets have been performed, however it is not clear whether rippling also produces jets found downstream of quasi-perpendicular shocks. We analyze four types of events in the quasi-perpendicular magnetosheath with signatures characteristic of magnetosheath jets, namely increased density and/or dynamic pressure, that were not produced by the shock rippling: 1) magnetic flux tubes connected to the quasi-parallel bow-shock, 2) non-reconnecting current sheets, 3) reconnection exhausts and 4) mirror mode waves. The flux tubes are downstream equivalents of the upstream traveling foreshocks. Magnetosheath jets can impact the magnetopause, so knowing the conditions under which they form may enable us to understand their signatures in the magnetosphere.
\end{abstract}

\section*{Plain Language Summary}
Magnetosheath jets have been identified as strong local enhancements of dynamic pressure or similar quantity in the magnetosheath associated to velocity and/or density enhancements. It is currently thought that $\sim$97~\% of magnetosheath jets form due to rippling of the quasi-parallel bow-shock. However shock rippling at the quasi-perpendicular shock occurs on much smaller spatial scales ($\sim$5~d$_i$, upstream ion inertial scales) than at quasi-parallel shock (several tens of d$_i$). It is thus not clear whether at the quasi-perpendicular shock the rippling produces magnetosheath jets.
Here we show for the first time four different phenomena, not associated to shock rippling, that can produce magnetosheath jets in the quasi-perpendicular magnetosheath. Three of them are already known types of events: current sheets, reconnection exhausts and mirror mode waves. The forth phenomena are magnetic flux tubes that are embedded in the quasi-perpendicular magnetosheath but are connected to the quasi-parallel bow-shock.

\section{Introduction}
The interaction of the solar wind (SW) with our planet's magnetosphere produces a supercritical \cite{treumann:2009} collisionless bow-shock upstream of Earth (e.g., \cite{tsurutani:1985}).
Depending on the angle between the local shock normal and the upstream B-field, $\theta_{BN}$, the bow-shock can be classified as quasi-parallel (Qpar, $\theta_{BN} <$ 45$^\circ$) or quasi-perpendicular (Qper, $\theta_{BN} >$ 45$^\circ$). Upstream of Qpar shocks the foreshock is formed \cite{eastwood:2005}. Such shocks present strong rippling at the spatial scales of $\lesssim$100 upstream ion inertial lengths (d$_i$, e.g., \cite{burgess:1989b, kraussvarban:1991}).

Downstream of the bow-shock there lies the magnetosheath (e.g. \cite{lucek:2005}) which is bounded by the magnetopause. Depending on which portion of the bow-shock the magnetosheath is magnetically connected to, we distinguish Qper and Qpar magnetosheath \cite{raptis:2020a}. The latter is populated by stronger B-field and plasma fluctuations and more energetic ions with energies of up to $\sim$30~keV.

Magnetosheath jets are also found in the magnetosheath (and the references therein \cite{plaschke:2018}).  \cite{hietala:2009} suggested that these jets form due to different processing of the SW at different locations on the Qpar bow-shock, caused by the rippling. \cite{hietala:2013} estimated that 97~\% of the observed jets are produced by the bow-shock ripples.
\cite{archer:2012} associated jets to IMF rotational discontinuities, while \cite{savin:2012} linked them to hot flow anomalies (e.g. \cite{lucek:2004b}.
\cite{karlsson:2012} associated a subset of magnetosheath jets, called plasmoids, to either plasmoids from the pristine SW or short large amplitude magnetic structures (e.g. \cite{giacolone:1993} from the foreshock that are transmitted into the magnetosheath.

In the past, statistical studies with large numbers of magnetosheath jets have been performed (\cite{plaschke:2013, plaschke:2016, archer:2013, raptis:2020b, raptis:2020a, liu:2020}), however their primary focus was not the jets' origin.

Knowing the causes of the jets is important, since it has been shown that they can perturb the geomagnetic field.  \cite{blancocano:2020} found jets formed by magnetic reconnection at the magnetopause. Some jets were observed to impact and sometimes penetrate the magnetopause \cite{hietala:2018, wang:2018, shue:2009, plaschke:2011, savin:2012, dmitriev:2012, plaschke:2016} and even perturb the ionosphere \cite{hietala:2012, archer:2013, wang:2018}. They were found by \cite{archer:2013} to be able to drive compressional and poloidal Pc5 (2-7~mHz) waves in the magnetosphere. Finally, their signatures in the data of ground-based magnetic observatories have been reported by \cite{dmitriev:2012} and \cite{archer:2013}.

In this work we show that certain phenomena in the Qper magnetosheath, some of them already known, may produce signatures in the spacecraft data, such as increased dynamic pressure (P$_{dyn}$), that would classify them as magnetosheath jets. These are magnetic flux tubes embedded in Qper magnetosheath that are connected to the Qpar bow-shock, non-reconnecting current sheets (CS), reconnection exhausts (RE) and mirror mode (MM) waves.

\section{Instrumentation}
We use data from three multi-spacecraft missions in orbit around Earth: Cluster \cite{escoubet:1997}, THEMIS \cite{angelopoulos:2008} and Magnetospheric Multiscale Mission (MMS, \cite{sharma:2005}). The Cluster probes carry several instruments, including a Fluxgate Magnetometer (FGM, \cite{balogh:2001})  and the Cluster Ion Spectrometer (CIS, \cite{reme:2001}). We use FGM B-field vectors and CIS-HIA ion moments with 0.2~s and 4~s time resolution, respectively. THEMIS data used in this work were provided by the Fluxgate Magnetometer (FGM, \cite{auster:2008}) and Ion Electrostatic Analyzer (iESA, \cite{mcfadden:2008}) with 0.25~s and 3~s resolution, respectively.
In the case of MMS we use B-field data that are provided by the Fluxgate Magnetometers (FGM, \cite{russell:2016}) with time resolution of 128 s$^{-1}$ and 16~s$^{-1}$ in the burst and survey mode, respectively. The ion data provided by the Fast Plasma Investigation (FPI, \cite{pollock:2016}) have 150~ms and 4~s time resolution in burst mode and survey mode, respectively.

\section{Observations}

In this section we show events that exhibited significantly enhanced P$_{dyn}$  in the Qper magnetosheath. These satisfy at least some of the criteria described in past literature, for example we selected them if they produced P$_{dyn}$ increases of $\geq$50~\% compared to the ambient values during a 10~minute time interval, as in \cite{gutynska:2015}. Additionally, two events, a passing flux tube and a CS, satisfy the criteria of \cite{archer:2012} that P$_{dyn}$ should exceed 1~nPa. The mirror-mode waves comply with the criteria of \cite{karlsson:2012} that the density increase inside the jets should increase by 50~\% compared to the surrounding values.

\subsection{Magnetic flux tubes connected to the Qpar bow-shock}

\begin{figure}
\centering
\noindent\includegraphics[width=0.8\textwidth]{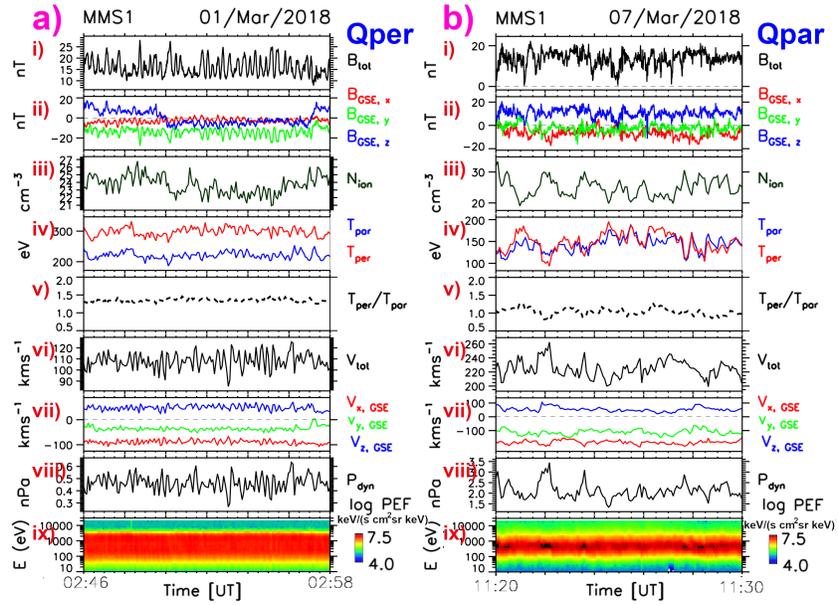}
\caption{MMS1 observations of the Qper (a) and Qpar (b) magnetosheath. The panels exhibit: i) B-field magnitude, ii) B-field components in the GSE coordinate system, iii) ion density, iv) ion parallel (blue) and perpendicular (red) temperatures, v) temperature anisotropy, vi) total ion velocity, vii) ion velocity components, viii) dynamic pressure and ix) ion spectrogram.}
\label{fig:threeMSH}
\end{figure}

We first discuss what is meant by the Qpar and Qper magnetosheath. Figure~\ref{fig:threeMSH} shows MMS1 observations during two time intervals, on 1 March 2018 (a) and on 7 March 2018 (b). The panels from i) to ix) exhibit: B-field magnitude and components in units of nT and in GSE coordinates (X-axis pointing from the Earth towards the Sun, the Y-axis lies in the ecliptic plane and is pointing towards dusk, Z-axis completes the right-hand system), the ion density (in cm$^{-3}$), parallel and perpendicular temperatures (eV), the temperature anisotropy defined as T$_{per}$/T$_{par}$, the total ion velocity and GSE components (kms$^{-1}$), the dynamic pressure (nPa) and the ion spectrogram with the colors representing the logarithm of the particle energy flux (PEF) in units of keV/(s cm$^2$ sr keV).

The crucial panels are those numbered with roman numbers iv), v) and ix). In the case of the Qper (Figure~\ref{fig:threeMSH}a) magnetosheath the ion T$_{per}$ (red) is larger than T$_{par}$ (blue). Thus, $T_{per}/T_{par}$ is above 1. Due to large T$_{per}$ the shocked SW signature in the ion spectrogram (red trace on panel ix) is very wide. Also, there is are very few ions at energies above 2500~eV. We compare these panels to those in Figure~\ref{fig:threeMSH}b) which exhibits the Qpar magnetosheath. Now T$_{par}$ and T$_{per}$ exhibit very similar values resulting in T$_{par}$/T$_{per}\sim$1. The values of the two temperatures are similar to  the T$_{par}$ in Figure~\ref{fig:threeMSH}a). Hence the red trace on the panel ix) is narrower, however the PEF intensity at energies above 2500~eV is much higher.

\begin{figure}
\noindent\includegraphics[width=1\textwidth]{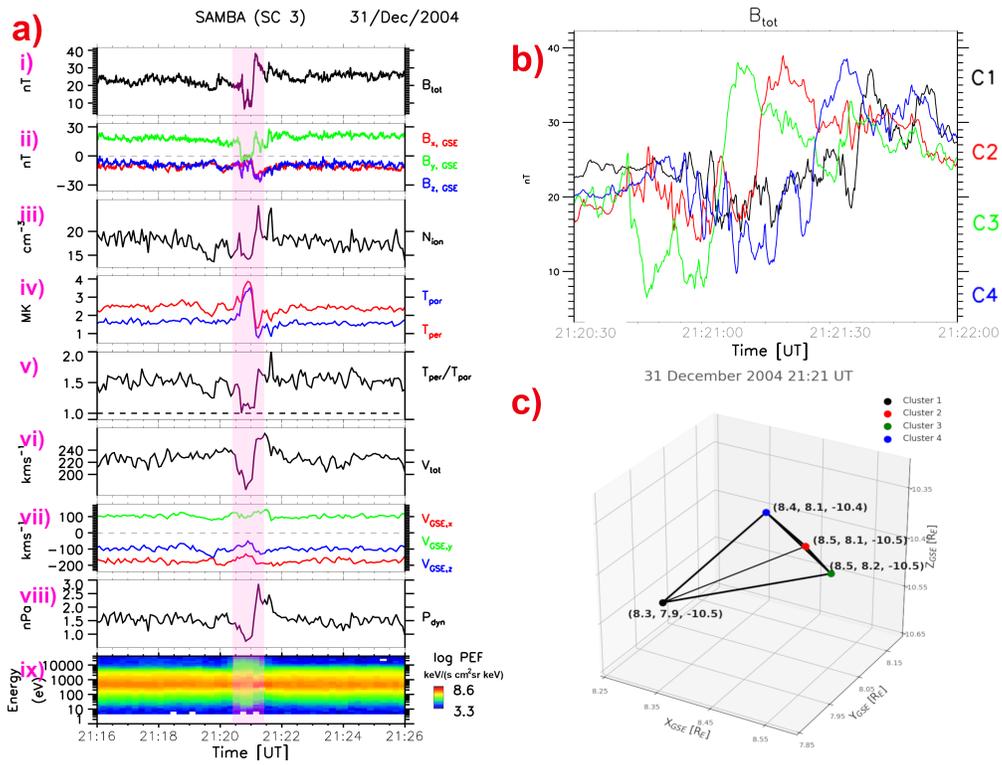}
\centering
\caption{a) Cluster~3 observations. The format is the same as in the Figure~\ref{fig:threeMSH}a). b) B-field profiles of the convected flux tube in (shaded region in a) in all four Cluster spacecraft data. c) Configuration of the Cluster constellation on 31 December 2004.}
\label{fig:FT311204}
\end{figure}

We now turn our attention to Figure~\ref{fig:FT311204}a). The format of this figure is basically the same as that of the Figure~\ref{fig:threeMSH}. During most of the exhibited time interval Cluster-3 is located in the Qper magnetosheath. However there is a short time period between 21:20:30-21:21:45~UT when ions with E$\lesssim$30~keV can be observed. During this time the B$_{tot}$, ion density, T$_{per}$/T$_{par}$ and total velocity values are diminished, while the T$_{per}$ and T$_{par}$ are enhanced. These signatures are very similar to those observed in the Qpar magnetosheath.

This region is bounded by rims where B$_{tot}$, N$_{ion}$ and V$_{tot}$ are enhanced and the temperature is diminished. In the upstream rim the B-field rotates during $\sim$4~s and this is also an approximate duration of this rim in the B$_{tot}$ data. The B-field rotation in the downstream rim is much longer, lasting $\sim$30~s. The high energy ions appear and disappear during these rotations. We can also observe that the P$_{dyn}$ in the downstream rim is strongly enhanced - it reaches 2.9~nPa, which represents a 107~\% increase compared to the ambient value of 1.4~nPa. It would be classified as encapsulated jet by \cite{raptis:2020b}.

This structure was convected pass the Cluster spacecraft. This can be seen in Figure~\ref{fig:FT311204}b) which shows B$_{tot}$ profiles from all Cluster probes. The first to detect the structure was C3, followed by C2, C4 and C1.
This order is the same as the order in which their X$_{GSE}$ coordinates decrease (Figure~\ref{fig:FT311204}c).

This structure is different from ``typical'' jets found in the Qpar magnetosheath. We argue here that the signatures featured in Figure~\ref{fig:FT311204} are due to a magnetic flux tube that was convected pass the Cluster probes. This tube exhibited either small transverse radius or it was wide and was crossed by the spacecraft near its edge.
Equivalent situations have been observed upstream of the Qper section of the Earth's bow-shock. Common phenomena there are the foreshock cavities (e.g. \cite{sibeck:2002, billingham:2008} which are observed in pristine SW, but the IMF and plasma properties inside them are the same as those typically found in the Earth's foreshock. Compared to their surroundings, foreshock cavities exhibit depressed density and magnetic field values in their cores, while these quantities are enchanced in their rims. Also, suprathermal ion fluxes are highly enhanced inside these events. \cite{omidi:2013} and \cite{kajdic:2017b} showed that foreshock cavities are a subset of traveling foreshocks with short duration in the spacecraft data. It was suggested that the traveling foreshocks form due to magnetic flux tubes that are observed upstream of the Qper bow-shock but are connected to the Qpar section of the bow-shock.

The reason we know that the observed event is caused by a magnetic flux tube and not by the back and forth motion of the boundary between the Qper/Qpar magnetosheath is because the convected flux tubes produce the so called convected signatures when they are observed by multiple spacecraft, while the back and forth boundary motion produces nested signatures (see \cite{kajdic:2017b}  for details). In the first case  (Figure~\ref{fig:FT311204}b) the sequence in which the spacecraft enter a passing magnetic flux tube is the same as the sequence in which they exit it. If the spacecraft had observed back and forth motion of the boundary between the Qper and Qpar magnetosheath, the Cluster probes would enter the Qpar magnetosheath in a certain sequence but the sequence in which they would exit it would be reversed.
Signatures of passing flux tubes have been reported in the past by, for example \cite{sibeck:2000} and \cite{katircioglu:2009} who studied their impact on the magnetopause motion. \cite{katircioglu:2009} called these events the magnetosheath cavities.

\begin{figure}
\centering
\noindent\includegraphics[width=1\textwidth]{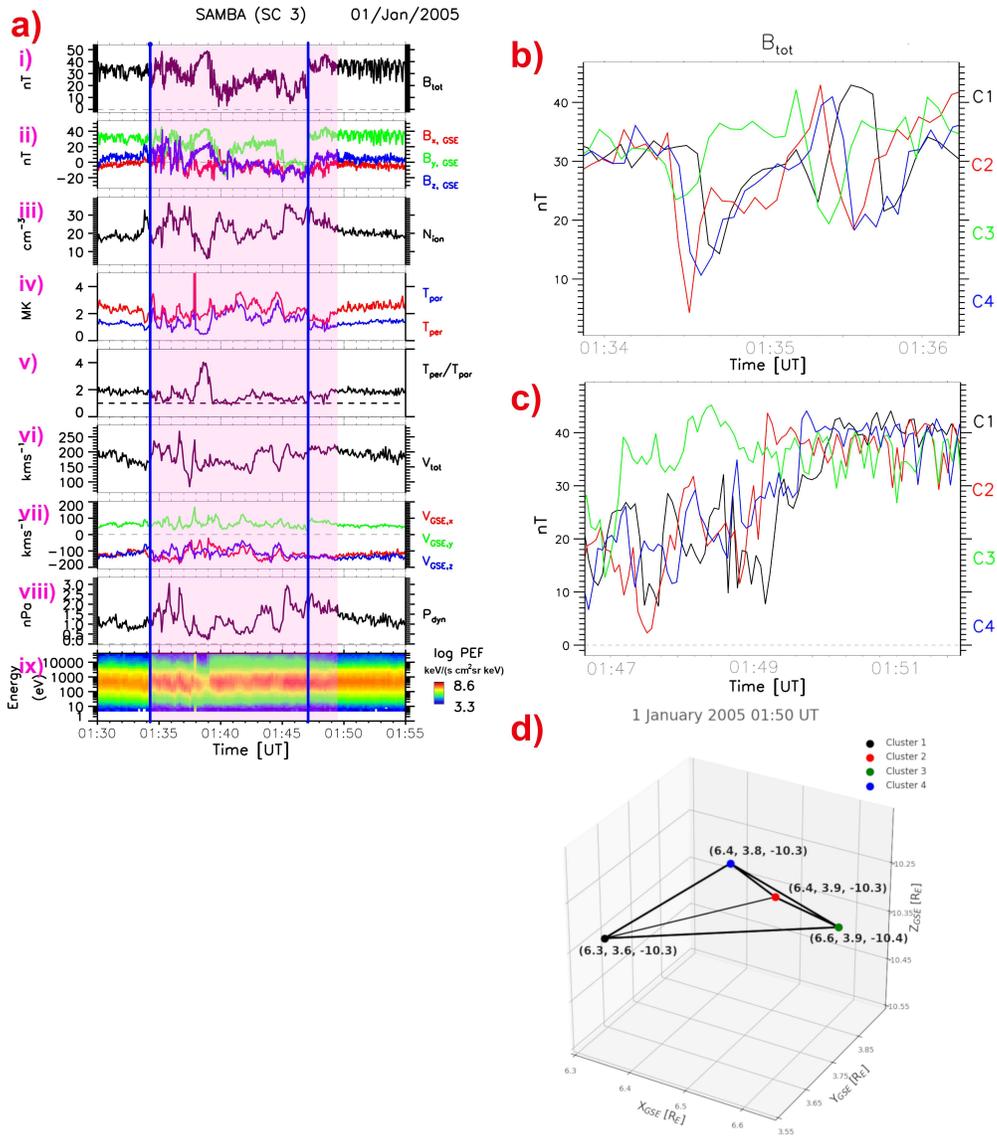}
\caption{a) Cluster~3 observations. The format is the same as in the Figure~\ref{fig:threeMSH}a) . b) and c) B-field profiles of the upstream and downstream edges of the flux tube in (left blue vertical lines in a) in all four Cluster spacecraft data. d) Configuration of the Cluster constellation on 1 January 2005.}
\label{fig:FT010105}
\end{figure}

In order to see what a longer lasting flux tube looks like in the data, we analyze the event in Figure~\ref{fig:FT010105}. We can observe IMF and plasma signatures of the Qper magnetosheath before and after the event (shaded in purple), while during the event IMF and plasma exhibit properties typical of those in the Qpar magnetosheath. The duration of this event in the data is $\sim$15~minutes. During the event the IMF and plasma parameters are highly perturbed producing several P$_{dyn}$ peaks (Qpar jets) with values of up to 3~nPa. These values are much higher than $\sim$1~nPa in the ambient Qper plasma. The reason for the P$_{dyn}$ peaks inside and at the rims of these flux tubes are beyond the scope of this paper. As before, the event is bounded by IMF rotations. We perform timing analysis of the event. In order to do so we found features that are recognizable in the data of all Cluster probes. We thus choose two short time intervals marked with vertical blue lines in Figure~\ref{fig:FT010105}a). These intervals are exhibited in Figures~\ref{fig:FT010105}b) and c). We can see that the C3 spacecraft was the first to observe the event and was followed by C2, C4 and C1. Since this order of the spacecraft is the same during both time intervals, we conclude that this structure was also convected pass the Cluster probes and is again a flux tube connected to the Qpar bow-shock.

\subsection{Non-reconnecting current sheet}
\begin{figure}
\centering
\noindent\includegraphics[width=1\textwidth]{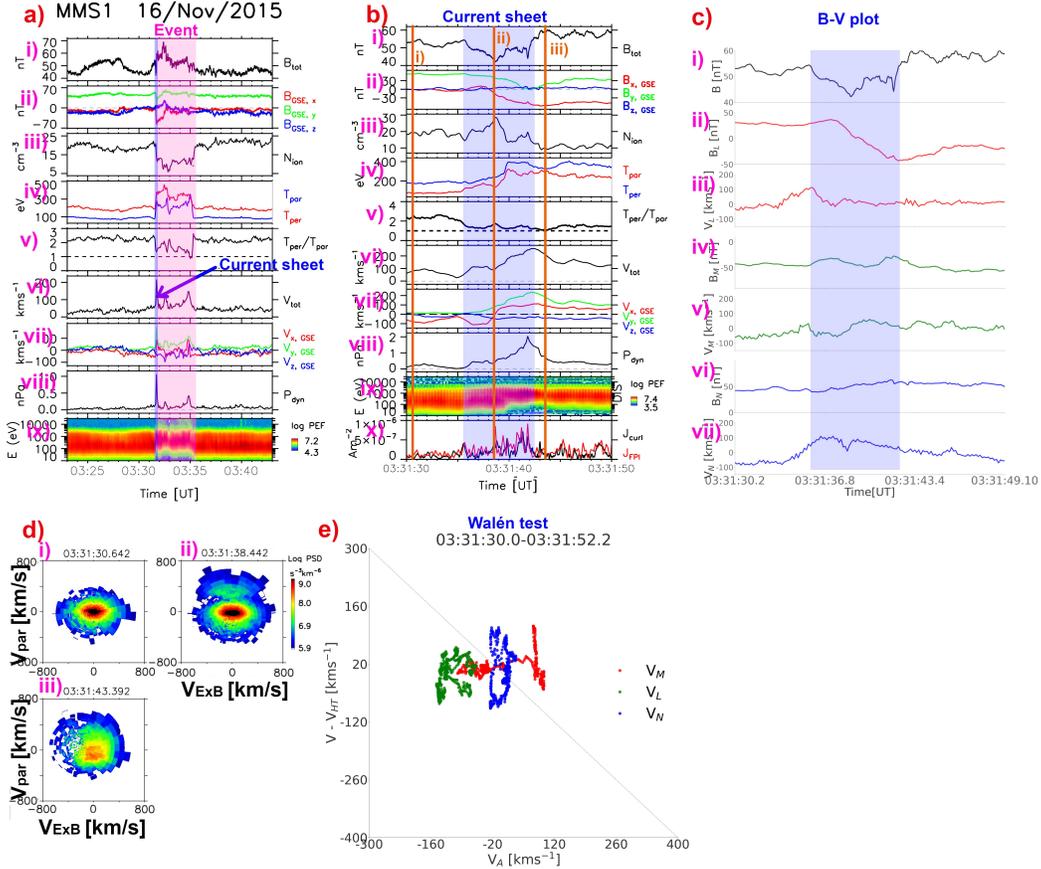}
\caption{MMS1 observations of a non-reconnecting current sheet (SC) on 16 November 2015. a) The event that caused the CS. b) A detailed view of the CS. Vertical lines mark the times of ion distribution in Figure~\ref{fig:figure4}d. b) B-V plot with i) IMF magnitude, (ii-vii) IMF and velocity in the N, L, M coordinates. d) Ion distributions before, during and after the CS. On $x$ ($y$) axis is the velocity perpendicular (parallel) to the B-field. d)  Wal\'en test. On $y$ axis is the ion velocity in the deHoffman-Teller frame and on $x$ axis is the Alfv\'en velocity. The data are in the LMN coordinates (e.g., \cite{voros:2017}.)}.
\label{fig:figure4}
\end{figure}

Figure~\ref{fig:figure4}a) features MMS1 observations during a 20 minute time interval on 16 November 2015. The spacecraft GSE coordinates were (10.5, 0.02, -0.54)~R$_E$. IMF and plasma parameters (in survey mode) indicate that during this time MMS1  was in the Qper magnetosheath except during 03:31:35-03:35:02~UT (shaded in purple), when a structure passed it. This event could be another flux tube albeit it is different from previous examples since it is less turbulent and the flux of ions with E$\lesssim$30~keV in it is quite low. Alternatively, similar events were identified by \cite{raptis:2020a} as possible flux transfer events  (\cite{paschmann:1982}) due to enhanced B-field magnitude, depleted densities, increased temperature, the presence of ions with E$\lesssim$30~keV and the southward pointing IMF at the time of the events (see \cite{petrinec:2020}). However, bipolar IMF signature is missing in our case. Finally, this event is similar to reconnection jets due to magnetopause reconnection \cite{blancocano:2020}, although these authors showed that such events exhibit ion distributions with two distinct populations, which is not the case here (Figure~\ref{fig:figure4}c iii). Hence, we will refer to this event simply as a ``structure''.

We now focus on a CS (Figure~\ref{fig:figure4}a, survey mode data) that produced a large P$_{dyn}$ peak ($\sim$1~nPa) at the upstream edge of the structure (vertical blue line). A short time interval exhibiting the CS in burst mode data is shown in Figure~\ref{fig:figure4}b). We can see that this in this data the P$_{dyn}$ peaks at $\sim$2~nPa. There is an additional panel panel x) that exhibits the electric current densities obtained by the curlometer method (black) and from particle moments (red). The CS exhibits some signatures reminiscent of reconnection exhausts, such as the drop of B$_{tot}$, increases of ion density and velocity and B-field and velocity rotations. In Figure~\ref{fig:figure4}d), which shows ion distributions before, during and after the CS, we see that during the CS there are two populations present - one from the ambient magnetosheath and the second one with velocity $>$200~kms$^{-1}$ parallel to the B-field. Although these ions might be accelerated in the CS itself, we note that ions with similar velocities in the plasma frame of reference also exist downstream of the CS, inside the ``structure'' (although there they appear at all angles with respect to the magnetic field) and could have simply leak from there into the CS. Further evidence against this being a reconnecting current sheet is presented in Figure~\ref{fig:figure4}c). There we show the IMF magnitude (panel i), B-field components and velocity components (ii-vii) in the NLM coordinates. These were obtained by performing the minimum variance analysis on the B-field data during the event. The velocity shown is a partial moment of the apparently accelerated component of the ion distribution function calculated for ion energies between $\sim$210~eV-3300~eV and pitch angles $\leq$45$^\circ$. We can see that B and V components do not show the required correlations at one edge and anti-correlations at the other one. Additionally, the Wal\'en test (see \cite{paschmann:2000} for details) in Figure~\ref{fig:figure4}e) also suggest that this event is not a reconnection exhaust, since the changes in B and V do not exhibit required (anti)correlations. Hence, we call this event a non-reconnecting CS.

The higher bulk velocity in the spacecraft rest frame is thus mainly due to the acceleration of the primary magnetosheath ion population there. A possible explanation for the observed high velocity and P$_{dyn}$ in the CS could be ion drifts caused by the magnetic field gradient and/or curvature across the CS.

\subsection{Reconnection exhaust}
Reconnection exhausts (RE) are ubiquitous in the pristine SW (e.g. \cite{gosling:2005a}. Although magnetic reconnection in the Qpar magnetosheath has been routinely observed by the MMS mission, ion jets originating from such regions, and thus REs, are not observed, except near the magnetopause. The lack of ion jets in the Qpar magnetosheath has been explained in terms of the high turbulence level there that does not permit their formation (e.g. \cite{phan:2018}. The Qper magnetosheath is much less turbulent and REs have been observed there (e.g. \cite{phan:2007a, oieroset:2017, eastwood:2018}.
Below we show an RE that produced large P$_{dyn}$ values.

\begin{figure}
\centering
\noindent\includegraphics[width=0.75\textwidth]{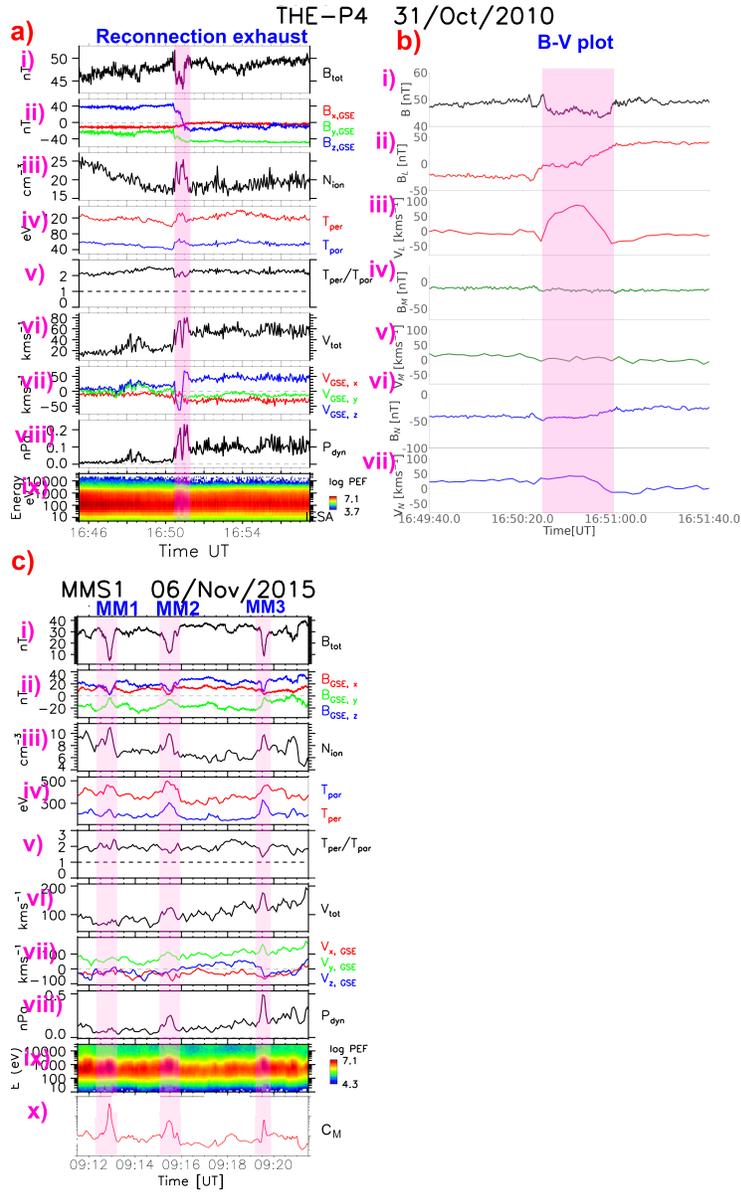}
\caption{a) THEMIS-E observations of a reconnection exhaust (shaded in purple) on 31 October 2010. The format of this figure is the same as that of the Figure~\ref{fig:FT311204}a). b) B-V plot with i) IMF magnitude, (ii-vii) IMF and velocity in the N, L, M coordinates, c) MMS1 observations of MM waves (shaded) on 6 November 2015.}
\label{fig:figure5}
\end{figure}

This event occured on 31 October 2010 (Figure~\ref{fig:figure5}) due to a single IMF discontinuity. It was observed by the THEMIS-E spacecraft and was studied previously in a different context by \cite{oieroset:2017}. At the time THEMIS-E was located at (11.0, -11.6, 2.4)~R$_E$ in GSE coordinates. The IMF vectors before and after the exhaust in GSE coordinates were (-15.2, -25.8, -6.0) and (-1.9, -52.3, 20.5), respectively, meaning that the IMF rotated by $\sim$43$^\circ$ across the exhaust. The event lasted for $\sim$32~seconds in the data (shaded in purple). We can see on panel viii) that this event produced a large P$_{dyn}$ increase. P$_{dyn}$ values before, during and after the RE were  0.01~nPa, 0.23~nPa and 0.1~nPa. Thus P$_{dyn}$ inside the RE was increased by 2200~\% and 130~\% with respect to the upstream and downstream regions, respectively. Figure~\ref{fig:figure5}b) shows B-field magnitude and GSE components of IMF and velocity. It can be observed that variations of B and V components are correlated on one side of the exhaust and anticorrelated on the other side, as it is expected for the REs.

\subsection{Mirror modes}
Mirror modes are ubiquitous in the magnetosheath (e.g. \cite{dimmock:2015b}. They exhibit compressive B-field fluctuations that are anticorrelated with the density and appear in the B-field data as dips or peaks. Magnetic dips are thus associated with density peaks and should produce P$_{dyn}$ enhancements.

Here we show MMS1 observations of MMs Figure~\ref{fig:figure5}c) during a 10~minute time interval on 6 November 2015, when the spacecraft were located at (11.0, 4.7, -0.8)~R$_E$ in GSE coordinates. An additional panel x) features the so called mirror parameter $C_M = \beta_i\perp\large(\frac{T_i\perp}{T_i\parallel}-1\large)$. This parameter was introduced by \cite{genot:2009}. The values of C$_M>1$ (C$_M<1$) denote plasma that is mirror-mode unstable (stable). We see that the plasma surrounding the MM waves is mirror-mode stable, while inside the MM waves it exhibits values of C$_M\gg$1.). Although the surrouding plasma is mirror-mode stable, it could have been unstable at earlier times when these waves were formed. Three most dominant MM waves are shaded in pink and marked as MM1, MM2 and MM3. These observations were made very near the magnetopause which was detected almost immediately after the featured time interval (not shown). In the B-field data the three mirror modes appear as dips that represent between 72~\% to 93~\% decrease compared to ambient values. They exhibit density and temperature enhancements. The temperature anisotropy is slightly increased in the case of MM1, it does not stand out from the ambient values ($\sim$1.8) for MM2 and is strongly, diminished in the case of MM3 (to $\sim$1.2, a 44~\% decrease).

The total velocity inside MM1 is not perturbed while it is increased inside MM2 (by $\sim$50~\%) and MM3 (by $\sim$116~\%) compared to their immediate neighborhood. The combination of density and velocity increases produces different signatures in the P$_{dyn}$ data. Compared to their immediate neighborhood the dynamic pressure inside these mirror modes is unchanged, increased by $\sim$270~\% and increased by  $\sim$250~\% inside MM1, MM2 and MM3, respectively. Due to the fact that the background P$_{dyn}$ values are highest around MM3, this structure stands out on panel viii). This event fulfills observational critera for a type of magnetosheath jets called diamagnetic plasmoids \cite{karlsson:2012, karlsson:2015} due to large density increase and B-field decrease.

\section{Conclusions}
In this work we show that magnetosheath jets in the Qper magnetosheath may have different origin than those in the Qpar magnetosheath. This is due to the fact that we do not expect to find magnetosheath jets produced at bow-shock ripples in the Qper magnetosheath.

We show that magnetosheath jet signatures can be produced by magnetic flux tubes that are embedded in the Qper magnetosheath but are connected to the Qpar section of the bow-shock. Inside them the IMF and plasma properties are the same as those typical of Qpar magnetosheath, namely these quantities are more turbulent and the temperature anisotropy drops to $\sim$1. Either rims and/or the insides of the flux tubes may produce P$_{dyn}$ peaks with values much higher than those in the surrounding Qper magnetosheath. We show that these flux tubes are convected pass the spacecraft by comparing their B-field profiles in the data of all Cluster probes. These flux tubes are the magnetosheath equivalent of foreshock cavities and traveling foreshocks commonly observed in the unperturbed upstream SW.% and to our best knowledge, this is the first time they are identified.

Next, we study a structure with a non-reconnecting CS at its upstream edge. This CS produced a magnetosheath jet with P$_{dyn}$ of 2~nPa, which, compared to the ambient value of 0.5~nPa, represents a $\sim$3000~\% increase.  We propose that the increased ion velocity and P$_{dyn}$ in the CS may be due to magnetic field gradient and/or curvature that produce ion drifts.

We further feature a reconnection exhaust in the Qper magnetosheath that occured due to single IMF discontinuity and is thus similar to those observed in the pristine SW (e.g. \cite{gosling:2005a}. During this event the P$_{dyn}$ is increased by 1200~\% and 270~\% compared to ambient values before and after the event, respectively.

Finally, we show that MMs can also present important P$_{dyn}$ enhancements. Two structures observed by MMS1 produced such peaks due to density and velocity enhancements inside them but only MM3 exhibited significant P$_{dyn}$ increase due to the fact that it also exhibited total velocity increase.

One of the reasons for which it is important to study magnetosheath jets is that they may cause perturbations of the geomagnetic field even at ground level (e.g. \cite{dmitriev:2012, archer:2013}.
Magnetosheath jets of different origin occur during different interplanetary conditions. Associating their signatures with jet formation mechanisms will undoubtedly be a topic of future investigations.

\section*{Acknowledgements}
The authors acknowledge the ClWeb team (http://clweb.irap.omp.eu/) for easy access to the data. Most of the figures were produced with ClWeb. The original datasets are available at the Cluster Science Archive (SCA, https://www.cosmos.esa.int/web/csa) and the Coordinated Data Analysis Web (CDAWeb, https://cdaweb.gsfc.nasa.gov/).
PK's work was supported by UNAM DGAPA/PAPIIT grant IN-105620. SR and TK are supported by Swedish National Space Agency (SNSA grant 90/17).
X.B.C.’s work was supported by UNAM DGAPA PAPIIT IN110921 grant.

%\bibliography{MyCitationLiterature}

\end{document}